\documentclass[a4paper]{jpconf}
\usepackage{graphicx}
\begin{document}
\title{Wide quantum critical region of valence fluctuations: 
\\
Origin of robust quantum criticality in quasicrystal Yb$_{15}$Al$_{34}$Au$_{51}$ under pressure
}

\author{Shinji Watanabe$^1$, Kazumasa Miyake$^2$}

\address{Department of Basic Sciences, Kyushu Institute of Technology, Kitakyushu, Japan$^1$ 
\\
Toyota Physical and Chemical Research Institute, Nagakute, Japan$^2$
}


\begin{abstract}
The mechanism of the emergence of robust quantum criticality in the heavy-electron quasicrystal Yb$_{15}$Al$_{34}$Au$_{51}$
is analyzed theoretically. By constructing a minimal model for the quasicrystal and its crystalline approximant, which contain concentric shell structures with Yb and Al-Au clusters, we show that a set of quantum critical points of the first-order valence transition of Yb appear as spots in the ground-state phase diagram. Their critical regions overlap each other, giving rise to a wide quantum critical region. This well explains the robust criticality observed in Yb$_{15}$Al$_{34}$Au$_{51}$ under pressure, 
and predicts the emergence of the common criticality in the crystalline approximant under pressure. 
The wider critical region in the quasicrystal than that in the crystalline approximant in the $T$-$P$ phase diagram and the field-induced valence-crossover ``region" in the $T$-$H$ phase diagram are predicted to appear. 
\end{abstract}

\section{Introduction}
Recently, a new heavy-electron metal Yb$_{15}$Al$_{34}$Au$_{51}$ with quasicrystal structure 
has been synthesized~\cite{Deguchi,Watanuki}. 
Intriguingly, it has been discovered that the quasicrystal Yb$_{15}$Al$_{34}$Au$_{51}$ shows unconventional quantum criticality 
such as the magnetic susceptibility $\chi\sim T^{-0.5}$, the NMR relaxation rate 
$(T_{1}T)^{-1}\propto\chi$, the electronic specific heat $C_{\rm e}/T\sim -\log{T}$, and approximately $T$-linear resistivity~\cite{Deguchi}.
This criticality is common to those observed in heavy-electron metals YbRh$_2$Si$_2$ and $\beta$-YbAlB$_4$ 
with periodic lattices, and is well explained by quantum criticality of Yb-valence fluctuations~\cite{WM2010} (see Table~\ref{table:QC}), which forms a new universality class different from 
that due to the antiferromagnetic criticality. 

\begin{table}
\caption{New type of quantum criticality in uniform magnetic susceptibility $\chi$, 
NMR/NQR relaxation rate $(T_{1}T)^{-1}$, specific-heat coefficient $C/T$, and 
resistivity $\rho$. As for valence criticality, $T>T_0$ regime with $T_0$ being 
characteristic temperature of critical valence fluctuation is shown (see \cite{WM2010} for detail). }
\label{tb:QC}
\begin{center}
\begin{tabular}{lcccc}
\hline
\multicolumn{1}{c}{Material/Theory} & \multicolumn{1}{c}{$\chi$} & \multicolumn{1}{c}{$(T_{1}T)^{-1}$} & 
\multicolumn{1}{c}{$C/T$} & \multicolumn{1}{c}{$\rho$} \\
\hline
Yb$_{15}$Al$_{34}$Au$_{51}$~\cite{Deguchi} & $T^{-0.51}$ & $\propto\chi$ & $-\ln{T}$ & $T$ \\
YbRh$_2$Si$_2$~\cite{Gegenwart} & $T^{-0.6}$ & $T^{-0.5}$ & $-\ln{T}$ & $T$ \\
$\beta$-YbAlB$_4$~\cite{Nakatsuji} & $T^{-0.5}$ & - & $-\ln{T}$ & $T^{1.5} \ \to \ T$  \\
Valence criticality~\cite{WM2010} & $T^{-0.5\sim-0.7}$ &  $\propto\chi$ & $-\ln{T}$ & $T$ \\
\hline
\label{table:QC}
\end{tabular}
\end{center}
\end{table}

Theoretically, it has been shown in \cite{WM2010} that 
the key origin of the emergence of the new type of quantum criticality is the locality of the critical Yb-valence fluctuation mode. 
This is ascribed to the atomic origin of the valence transition at the Yb site, 
and is considered to be not depending on the detail of lattice structures 
such as the periodic lattice or quasicrystal.

Interestingly, in Yb$_{15}$Al$_{34}$Au$_{51}$ quantum criticality emerges without applying pressure and magnetic field, and the criticality is quite robust against applying hydrostatic pressure~\cite{Deguchi}. To clarify the mechanism of the zero-tuning quantum criticality as well as the robust criticality, we have theoretically analyzed electronic states of quasicrystal Yb$_{15}$Al$_{34}$Au$_{51}$ and its crystalline approximant~\cite{WM2013}. 
Since the detailed results were reported in \cite{WM2013}, here we outline the key points and discuss the consequence of 
the theoretical results, which will be compared with experiments in the quasicrystal and crystalline approximant. 

\section{Analysis of the Yb-Al-Au cluster}

\begin{figure}[b]
\includegraphics[width=14.0cm]{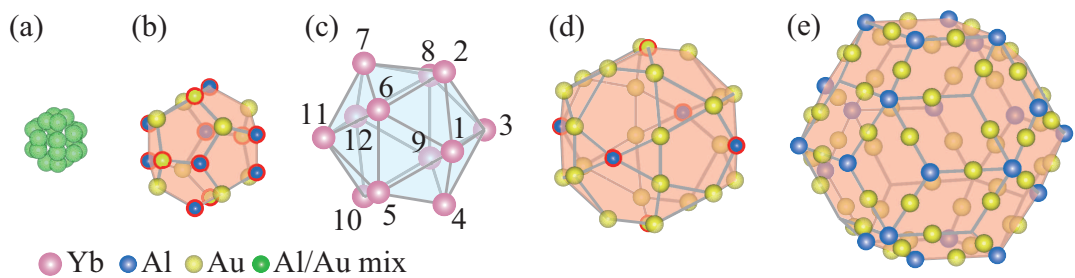}
\caption{(color online) Concentric shell structures of Tsai-type cluster in the Yb-Al-Au approximant:
(a) first shell, (b) second shell, (c) third shell, (d) fourth shell, and (e) fifth shell. 
The number in (c) indicates the $i$-th Yb site.
}
\label{fig:Yb_cluster}
\end{figure}

Figure~\ref{fig:Yb_cluster} shows concentric shell structures of Tsai-type cluster 
in the Yb-Al-Au approximant, which is the basic structure of 
the quasicrystal Yb$_{15}$Al$_{34}$Au$_{51}$~\cite{Deguchi}. 
The approximant has periodic arrangement of the body-centered cubic structure whose unit cell 
contains the shell structures shown in Fig.~\ref{fig:Yb_cluster}(a)-(e). 
In the second shell, 12 sites are the Al/Au mixed sites 
(the sites framed in red in Fig.~\ref{fig:Yb_cluster}(b)) 
where the Al or Au atom exists  
with the rate of $62~\%/38~\%$, respectively~\cite{Ishimasa}. 
In the fourth shell, 6 sites are the Al/Au mixed sites 
(the sites framed in red in Fig.~\ref{fig:Yb_cluster}(d))  
where the Al or Au atom exists 
with the rate of $59~\%/41~\%$, respectively~\cite{Ishimasa}.
These rates are average values of the whole crystal. 
Hence the location of the Al and Au sites and the existence ratio can be 
different at the next-to-next concentric shells each other both in the quasicrystal and approximant. 
Thus the 2nd and 4th shells illustrated in Figs.~\ref{fig:Yb_cluster}(b) and 1(d), 
respectively, are such examples. 

As shown in \cite{WM2010}, 
almost dispersionless critical valence-fluctuation mode is the key origin of the emergence 
of the new type of quantum criticality shown in Table~\ref{table:QC}. 
This implies that locality of valence fluctuation is essentially important. 
Namely, charge transfer between the Yb site and surrounding atoms, which is basically local, 
is considered to play a key role. 
Hence, we concentrate on the Yb-Al-Au cluster shown in Fig.~\ref{fig:Yb_cluster}.

Recently, the same measurement as in Table~\ref{tb:QC} performed by replacing Al with Ga has revealed that 
quantum critical behavior in the physical quantities disappears~\cite{Matsukawa}. 
This suggests that conduction electrons at the Al site contribute to 
the quantum critical state.  
Hence, we consider the simplest minimal model for the 4f-hole orbital at the Yb site and 
conduction-hole orbital at the Al site on the Yb-Al-Au cluster:  
\begin{eqnarray}
H&=&-\sum_{\alpha=2,5}\sum_{\langle \xi\nu\rangle\sigma}^{\rm (\alpha)}t_{\xi\nu}^{(\alpha)}\left(
c_{\xi\sigma}^{\dagger}c_{\nu\sigma}+{\rm h.c.}
\right)
-\sum_{\xi\sigma}^{(4)}\sum_{\eta}^{\rm (5)}t_{\xi\eta}\left(
c_{\xi\sigma}^{\dagger}c_{\eta\sigma}+{\rm h.c.}
\right)
\nonumber
\\
&+&\varepsilon_{\rm f}\sum_{i=1 \sigma}^{12}n_{i\sigma}^{\rm f}
+U\sum_{i=1}^{12}n_{i\uparrow}^{\rm f}n_{i\downarrow}^{\rm f}
+\sum_{i=1\sigma}^{12}\sum_{\eta}^{\rm (2,4,5)}V_{i\eta}
\left(
f_{i\sigma}^{\dagger}c_{\eta\sigma}+{\rm h.c.}
\right)
\nonumber
\\
&+&U_{\rm fc}\sum_{i=1\sigma}^{12}\sum_{\eta\sigma'}^{\rm (2,4)}
n_{i\sigma}^{\rm f}n_{\eta\sigma'}^{\rm c}
\label{eq:PAM}
\end{eqnarray}
where $f_{j\sigma}^{\dagger}$ $(f_{j\sigma})$ and $c_{j\sigma}^{\dagger}$ $(c_{j\sigma})$ 
are a creation (annihilation) operator of the f hole and 
the conduction hole at the $j$-th site with spin $\sigma$, respectively, and 
$n^{\rm f}_{j\sigma}\equiv f_{j\sigma}^{\dagger}f_{j\sigma}$ 
and $n_{j\sigma}^{\rm c}\equiv c_{j\sigma}^{\dagger}c_{j\sigma}$. 
The first term represents the conduction-hole transfer on the 2nd and 5th shells, 
where 
$\sum_{\langle \xi\nu\rangle}^{(\alpha)}$ denotes the summation of the nearest-neighbor Al sites 
on the $\alpha$-th shell ($\alpha$=2: Fig~\ref{fig:Yb_cluster}(b) and $\alpha$=5: Fig.~\ref{fig:Yb_cluster}(e)). 
The second term represents the conduction-hole transfer between the 4th and 5th shells, 
where $\sum_{\xi\sigma}^{(4)}\sum_{\eta}^{(5)}$ denotes the summation of the nearest-neighbor Al sites on the 5th shell 
for each Al site on the 4th shell (Fig.~\ref{fig:Yb_cluster}(d)). 
The third and fourth terms represent the f-hole energy level $\varepsilon_{\rm f}$ and 
onsite Coulomb repulsion $U$ on the 3rd shell (Fig.~\ref{fig:Yb_cluster}(c)), respectively. 
The fifth term represents the hybridization $V_{i\eta}$ between f and conduction holes,  
where 
$\sum_{i=1\sigma}^{12}\sum_{\eta}^{(\alpha)}$ denotes the summation of the nearest-neighbor sites 
on the $\alpha$-th shell ($\alpha$=2, 4, or 5) for each $i$-th Yb site on the 3rd shell.  
The last term represents the inter-orbital Coulomb repulsion $U_{\rm fc}$.
This term has been shown theoretically to be essentially important to cause the quantum criticality of 
Yb-valence fluctuations~\cite{WM2010}. 
Namely, this model has essentially the same structure as the periodic Anderson model with the $U_{\rm fc}$ term which exhibits the quantum valence criticality shown in Table~\ref{table:QC}.

\begin{figure}[h]
\includegraphics[width=14cm]{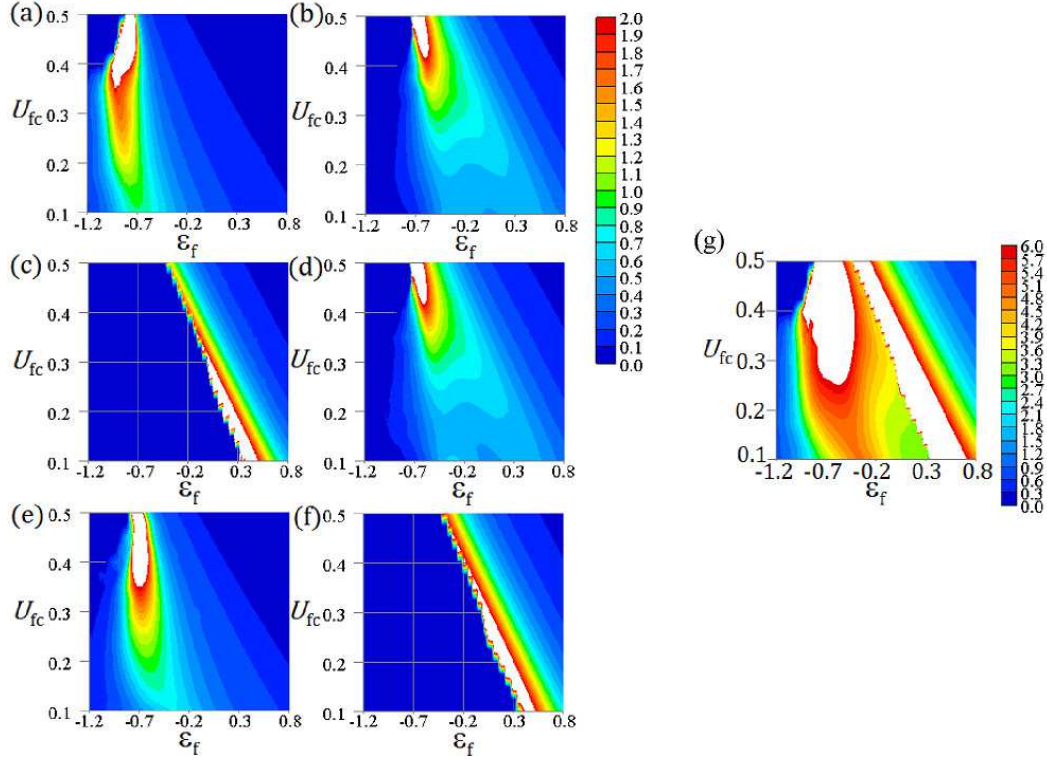}
\caption{(color online) Contour plot of valence susceptibility $\chi_{{\rm v}i}$ 
for (a) $i$=1, (b) 2, (c) 3, (d) 4, (e) 5, and (f) 6 
in the $\varepsilon_{\rm f}$-$U_{\rm fc}$ plane. 
Each white region shows  
diverging critical valence fluctuation arising from the VQCP for each $i$. 
(g) Contour plot of total valence susceptibility $\chi_{\rm v}=\sum_{i=1}^{12}\chi_{{\rm v}i}$ 
in the $\varepsilon_{\rm f}$-$U_{\rm fc}$ plane.
White regions represent diverging critical valence fluctuations arising from the VQCP's 
for $i=1\sim 12$. 
Note that contour values are shown in larger scale than that in (a)-(f).
}
\label{fig:chiv1_6}
\end{figure}

To understand the fundamental nature of this system, 
here we set $t_{\xi\nu}^{(2)}=t_{\xi\nu}^{(5)}=t_{\xi\eta}=t=1$, $V_{i\eta}=V=0.3$, and $U=\infty$ 
for a typical parameter set of heavy-electron systems. 
We consider the case that the total hole number   
$(N_{\uparrow},N_{\downarrow})=(24,24)$ in the Yb-Al-Au cluster with $N=54$ Yb and Al atoms in total, 
as shown in Fig.~\ref{fig:Yb_cluster}. 
By using the slave-boson mean-field theory~\cite{Read1983}, 
we have calculated the $\varepsilon_{\rm f}$ dependence of 
$\langle n_{i}^{\rm f}\rangle=\langle n_{i\uparrow}^{\rm f}\rangle +\langle n_{i\downarrow}^{\rm f}\rangle$ 
for each $U_{\rm fc}$ in the ground state 
and obtained the valence susceptibility defined by 
$\chi_{{\rm v}i}\equiv -\partial\langle n_{i}^{\rm f}\rangle/\partial\varepsilon_{\rm f}$, 
as shown in Figs.~\ref{fig:chiv1_6}(a)-(f).

When the f level is located at a deep position, i.e., 
$\varepsilon_{\rm f}$ is small enough in eq.~(\ref{eq:PAM}), 
one f hole is located at the $i$-th site on the 3rd shell (see Fig.~\ref{fig:Yb_cluster}(c)) with $\langle n^{\rm f}_{i}\rangle=1$, 
as a result of on-site strong hole correlations caused by $U=\infty$. 
As $U_{\rm fc}$ increases, at the point which satisfies 
%
$
\varepsilon_{\rm f}+U_{\rm fc}\sum_{\eta}^{(2,4)}\langle n_{\eta}^{\rm c}\rangle\approx \mu, 
$
%
where $\mu$ is the chemical potential and $\eta$ specifies the conduction-hole orbital corresponding to the $i$-th Yb site~\cite{note}, $\langle n_{i}^{\rm f}\rangle$ shows a jump to 
the smaller $\langle n_{i}^{\rm f}\rangle$ value. 
Namely, the first-order valence transition (FOVT) occurs 
since large $U_{\rm fc}$ forces holes pour into either the f level or the conduction orbital. 
As $\varepsilon_{\rm f}$ increases or $U_{\rm fc}$ decreases along the FOVT line, 
the value of the valence jump 
decreases and finally disappears at the quantum critical end point of the FOVT. 
This point is called 
the quantum critical point of the valence transition (VQCP),  
at which the critical valence fluctuation diverges, i.e., $\chi_{{\rm v}i}=\infty$.
As further $\varepsilon_{\rm f}$ increases or $U_{\rm fc}$ decreases along the valence-crossover line 
extended from the FOVT line, the valence susceptibility $\chi_{{\rm v}i}$ is still enhanced, 
giving rise to the quantum critical region in the $\varepsilon_{\rm f}$-$U_{\rm fc}$ plane.  
Since the hybridization paths between each $i$-th Yb site and surrounding the nearest-neighbor Al sites 
are different owing to the presence of the Al/Au mixed sites shown in Fig.~\ref{fig:Yb_cluster}, 
the location of the VQCP for each $i$ becomes different. 
Since we now consider the case that 
Al atoms are located at the Al/Au mixed sites in the 2nd shell and 4th shell 
with inversion symmetry with respect to the cluster center as shown 
in Figs.~\ref{fig:Yb_cluster}(b) and \ref{fig:Yb_cluster}(d), respectively, 
the results for the $i=1\sim 6$-th Yb sites in Figs.~\ref{fig:chiv1_6}(a)-(f), respectively, 
are the same as those for the corresponding $i=7\sim 12$-th Yb sites with inversion symmetry. 


An important result here is that the VQCP's appear as spots in the ground-state phase diagram, 
whose critical regions are overlapped to be unified, giving rise to a wide quantum-critical-region.
Actually, the total valence susceptibility $\chi=\sum_{i=1}^{12}\chi_{{\rm v}i}$, 
corresponding to experimental observation of 
quantum criticality, shows that the wide critical region appears 
in the phase diagram (see~Fig.~\ref{fig:chiv1_6}(g)). 
Note that although we now consider the case with inversion symmetry on the Al sites 
(see Fig.~\ref{fig:Yb_cluster}), absence of the symmetry (expected in real quasicrystal structure)n gives rise to 12 VQCP spots 
but not 6 spots per an Yb-Al-Au cluster. 
This makes the critical region be further enlarged in the $\varepsilon_{\rm f}$-$U_{\rm fc}$ phase diagram.

\section{Consequence of Theoretical Results and its Relevance to Quasicrystal and Crystalline Approximant}

We have analyzed the fundamental nature of the Yb-Al-Au cluster 
which is the core structure of both the quasi-crystal and crystalline approximant. 
By further considering outer concentric shells to the Yb-Al-Au cluster, 
the quasicrystal structure is constructed, while by considering periodic arrangement of the concentric shells 
in Fig.~\ref{fig:Yb_cluster} 
as a unit cell, the crystalline approximant is constructed. 
Although comparison of the electronic states in the bulk limit of both systems should be made 
for complete understanding of each system, the fundamental properties clarified in the previous section 
are considered to be preserved even in the bulk limit 
since the critical valence fluctuation is ascribed to the atomic origin so that the locality is essential. 

When we apply pressure to Yb-based compounds, the f-hole level decreases 
and $U_{\rm fc}$ increases in general. 
Hence, in case that applying pressure follows the line located in the enhanced 
critical-valence-fluctuation region, robust quantum criticality appears 
under pressure. 
The emergence of a wide critical region gives a natural explanation for why 
quantum criticality appears without fine tuning of control parameters such as pressure and magnetic field in Yb$_{15}$Al$_{34}$Au$_{51}$. 

Essentially local nature of critical valence fluctuations offers an 
interesting possibility that the same criticality as that of the quasicrystal 
appears in the crystalline approximant when pressure is tuned. 
Namely, the quasicrystal is considered to be located at the enhanced critical valence-fluctuation region 
in the ground-state phase diagram 
as noted above, while the crystalline approximant seems slightly away from it. 
Since $\chi$, $(T_1T)^{-1}$, and $C_{\rm e}/T$ for $T\to 0$ in the crystalline approximant 
are being suppressed in comparison with those of 
the quasicrystal at ambient pressure~\cite{Deguchi},  
applying hydrostatic pressure to the crystalline approximant is expected to make the system 
approach the enhanced critical valence-fluctuation region. 

In the quasicrystal constructed by arranging the outer concentric shells to the Yb-Al-Au cluster shown in Fig.~\ref{fig:Yb_cluster},   
the VQCP's are expected to appear as 
widespread and condensed spots (like, for example, the Andromeda Galaxy). 
On the other hand, in the crystalline approximant, the Yb-12 cluster is 
periodically arranged, giving rise to in principle 12 VQCP spots in the bulk limit. 
Hence, the quantum critical region in the quasicrystal is expected to be wider than that 
in the crystalline approximant. 
The recent measurement has revealed that the pressurized crystalline approximant exhibits the same criticality as that of the quasicrystal with its critical region being smaller than that in the $T$-$P$ phase diagram of the quasicrystal~\cite{Deguchi_SCES2014}.  
The emergence of the common criticality in both the quasicrystal and crystalline approximant strongly suggests that 
the key origin of the quantum criticality arises from the Yb-Al-Au cluster with  
the locality of critical Yb-valence fluctuations as the underlying mechanism, as emphasized above. 
The importance of the Yb-Al-Au cluster is also reinforced by the experimental fact that the simple Kondo-disorder scenario 
due to the Al/Au mixed sites is incompatible with the robustness of the critical exponent under pressure~\cite{Deguchi}. 

\begin{figure}[h]
\includegraphics[width=9cm]{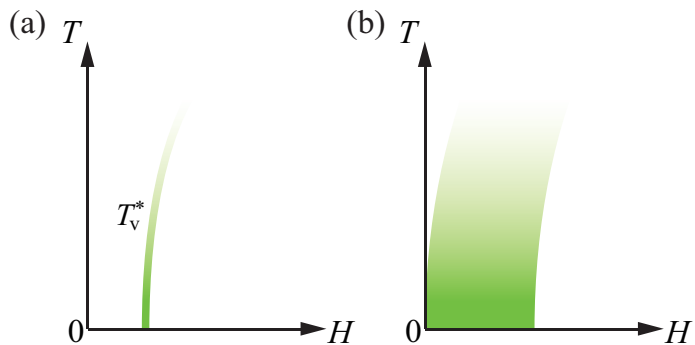}
\caption{\label{fig:Tv_H}(color online) (a) Filed-induced valence-crossover line $T_{\rm v}^{*}$ in the $T$-$H$ phase diagram 
of the periodic-lattice system.  
(b) Field-induced valence-crossover ``region" in the $T$-$H$ phase diagram of the quasycrystal.}
\end{figure}

In the periodic-lattice system, the periodic Anderson model with the Coulomb repulsion $U_{\rm fc}$ between f and conduction orbitals, 
which has essentially the same structure as the present model given by eq.~(\ref{eq:PAM}), was theoretically studied under a magnetic field~\cite{WM2008}. 
When we apply a magnetic field, the VQCP starts to shift to the smaller-$U_{\rm fc}$ direction in the $\varepsilon_{\rm f}$-$U_{\rm fc}$ phase diagram~\cite{WM2008}. 
If the system is located at the Kondo regime near the VQCP, the field-induced valence-crossover line $T_{\rm v}^{*}(H)$, $H$ being a magnetic field, appears, 
as shown in Fig.~\ref{fig:Tv_H}(a). 
Experimentally, such a behavior was observed in the heavy-electron metal YbAuCu$_4$ by the $^{63}$Cu-NQR frequency change under a magnetic field. Note that YbAuCu$_4$ is a sister compound of the prototypical valence-transition material YbInCu$_4$~\cite{Wada_2008}. 
As $H$ goes across the $T_{\rm v}^{*}(H)$ line at the fixed low $T$ in Fig.~\ref{fig:Tv_H}(a), the change from the Kondo regime with the relatively-larger Yb valence to the mixed-valence regime with the relatively-smaller Yb valence occurs in addition to the usual Zeeman effect where the Yb valence monotonically increases 
 (for detail, see section~4.2 in \cite{WM2011}). 
Our result shown in Fig.~\ref{fig:chiv1_6}(g) suggests that, 
in the quasicrystal Yb$_{15}$Al$_{34}$Au$_{51}$, the VQCP's are densely condensed, 
giving rise to the valence-crossover ``region" with a certain width of a magnetic field, 
as shown in Fig.~\ref{fig:Tv_H}(b),  
where the negative contribution to monotonic increase in the Yb valence due to the Zeeman effect is expected to be observed. 
The negative contribution is expected to be remarkable at lower temperatures since the valence-crossover line arises from the 
critical end point of the FOVT which is considered to be located in the vicinity of $T=0$~K or 
even at the negative-$T$ side~\cite{WM2008,WM2011}. 
Although the magnitudes of the negative and positive contributions in the Yb valence depend on the detail of the material, 
suppression of the monotonic increase in the Yb valence for a certain field range at low temperatures  
is expected to be observed in Yb$_{15}$Al$_{34}$Au$_{51}$. 
Detection of such a nontrivial negative contribution to the Yb-valence increase under a magnetic field is an interesting subject in future studies.

\section{Summary}
We have discussed that the Yb-Al-Au cluster with critical Yb-valence fluctuations 
plays a key role in the emergence of 
the robust quantum criticality in the quasicrystal Yb$_{15}$Al$_{34}$Au$_{51}$ under pressure. 
Our analysis of the Yb-Al-Au cluster predicts the emergence of the common criticality even 
in the crystalline approximant under pressure whose critical region is expected to be smaller than that 
of the quasicrystal in the $T$-$P$ phase diagram,  
and the emergence of the field-induced Yb-valence crossover ``region" in the $T$-$H$ phase diagram. 

\ack
This work is supported by Grants-in-Aid for Scientific Research (No. 24540378 and No. 25400369) from Japan Society for the Promotion of Science (JSPS). 
One of us (S.W.) is supported by JASRI (Proposal 
Nos. 0046 in 2012B, 2013A, 2013B, and 2014A).


\begin{thebibliography}{99}
\bibitem{note} Note that the second term in this equation has different values for each $i$.
\end{thebibliography}
\end{document}